\documentclass[prl,twocolumn,english,superscriptaddress]{revtex4-1}

\usepackage[T1]{fontenc}
\usepackage[latin9]{inputenc}
\usepackage{color}
\usepackage{babel}
\usepackage{latexsym}
\usepackage{float}
\usepackage{amsmath}
\usepackage{amsfonts}
\usepackage{graphicx}
\usepackage{times}   
\usepackage{esint}
\usepackage{subfigure}
\usepackage{verbatim}
\usepackage{footmisc}
\usepackage[unicode=true,pdfusetitle,
 bookmarks=false,colorlinks=true,citecolor=blue,urlcolor=blue,linkcolor=red]{hyperref}
\makeatletter

\@ifundefined{textcolor}{}
{%
 \definecolor{BLACK}{gray}{0}
 \definecolor{WHITE}{gray}{1}
 \definecolor{RED}{rgb}{1,0,0}
 \definecolor{GREEN}{rgb}{0,1,0}
 \definecolor{BLUE}{rgb}{0,0,1}
 \definecolor{CYAN}{cmyk}{1,0,0,0}
 \definecolor{MAGENTA}{cmyk}{0,1,0,0}
 \definecolor{YELLOW}{cmyk}{0,0,1,0}
}

\@ifundefined{date}{}{\date{}}
\AtBeginDocument{
  
}
\makeatother

\newcommand{\SAVE}[1]{}

\newcommand{\prlsec}[1]{\emph{#1---}}

\begin{document}
\renewcommand{\thefootnote}{\fnsymbol{footnote}}
\renewcommand\abstractname{}
\title{The mother of all states of the kagome quantum antiferromagnet}

\author{Hitesh J. Changlani}
\affiliation{Department of Physics and Astronomy, Johns Hopkins University, Baltimore, Maryland 21218, USA}
\affiliation{Institute for Quantum Matter, Johns Hopkins University, Baltimore, Maryland 21218, USA}
\affiliation{Department of Physics and Institute for Condensed Matter Theory, University of Illinois at Urbana-Champaign, 
1110 West Green St, Urbana IL 61801, USA}
\author{Dmitrii Kochkov}
\affiliation{Department of Physics and Institute for Condensed Matter Theory, University of Illinois at Urbana-Champaign, 
1110 West Green St, Urbana IL 61801, USA}
\author{Krishna Kumar}
\affiliation{Department of Physics and Institute for Condensed Matter Theory, University of Illinois at Urbana-Champaign, 
1110 West Green St, Urbana IL 61801, USA}
\author{Bryan K. Clark}
\affiliation{Department of Physics and Institute for Condensed Matter Theory, University of Illinois at Urbana-Champaign, 
1110 West Green St, Urbana IL 61801, USA}
\author{Eduardo Fradkin}
\affiliation{Department of Physics and Institute for Condensed Matter Theory, University of Illinois at Urbana-Champaign, 
1110 West Green St, Urbana IL 61801, USA}
\date{\today}

\begin{abstract}
Frustrated quantum magnets are a central theme in condensed matter physics due to the richness of their phase diagrams.  
They support a panoply of phases including various ordered states and topological phases. 
Yet, this problem has defied solution for a long time due to the lack of controlled approximations which 
make it difficult to distinguish between competing phases. Here we report the discovery of a special \emph{quantum} macroscopically 
degenerate point in the $XXZ$ model on the spin 1/2 kagome quantum antiferromagnet for the ratio of Ising to antiferromagnetic transverse coupling $J_z/J=-1/2$. 
This point is proximate to many competing phases explaining the source 
of the complexity of the phase diagram. 
We identify five phases near this point including both spin-liquid and broken-symmetry phases and give evidence that the kagome Heisenberg antiferromagnet is close to a transition between two phases. 
\end{abstract}

\maketitle
The history of quantum frustrated magnetism began in 1973 with 
Anderson's suggestion that the ground state of the nearest-neighbor (n.n.) Heisenberg model on the triangular lattice 
was a quantum spin-liquid~\cite{Anderson73}.  While we now know that this particular model 
does not support a spin-liquid, both experimental and theoretical evidence 
has been building for quantum spin-liquids in various lattices built of triangular motifs. 
Materials such as Herbertsmithite (a kagome lattice of Cu$^{2+}$ ions)~\cite{Nocera_Kagome_2007} 
and $\textrm{Na}_4\textrm{Ir}_3\textrm{O}_8$ (a hyper-kagome lattice of Ir$^{4+}$ ions)~\cite{Hyperkagome_QSL} 
fail to order down to low temperatures suggesting a possible spin-liquid ground state. 
This is supported by theoretical calculations which show that a panoply of spin-liquids (or exotic ordered phases) occur 
in a variety of Hamiltonians~\cite{ZengElser, RRPSingh_Huse, Wen_Kagome, White_Kagome, Depenbrock_Kagome, Iqbal_Kagome, Jiang_Balents, 
Clark_kagome, Tay_Motrunich, He_Zaletel_Kagome, Normand_Xiang, YCHeXXZ, Changlani_Lauchli, Yao_Zaletel}. 
This Letter presents an explanation of multiple energetically competitive phases in these models. 
\begin{figure}
\centering
\includegraphics[width=0.495\linewidth]{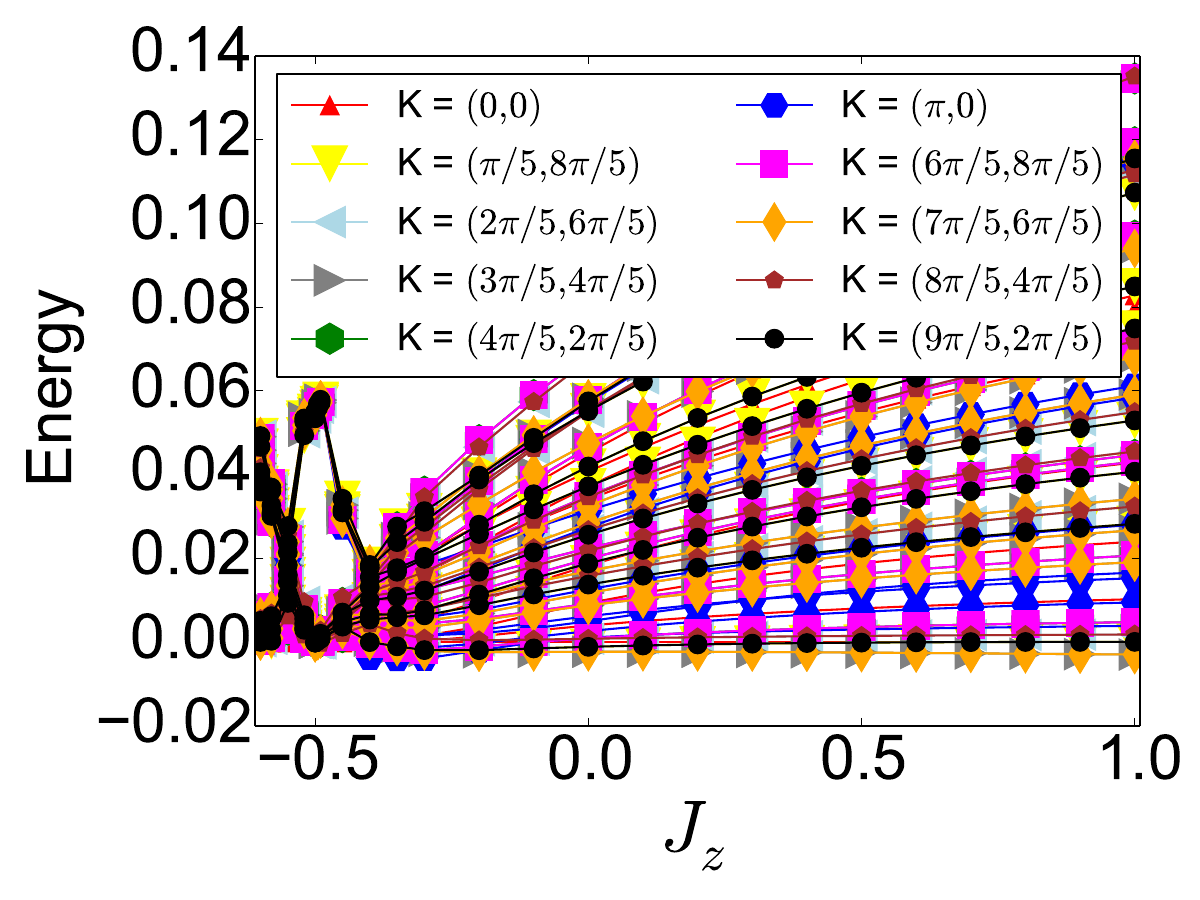}
\includegraphics[width=0.495\linewidth]{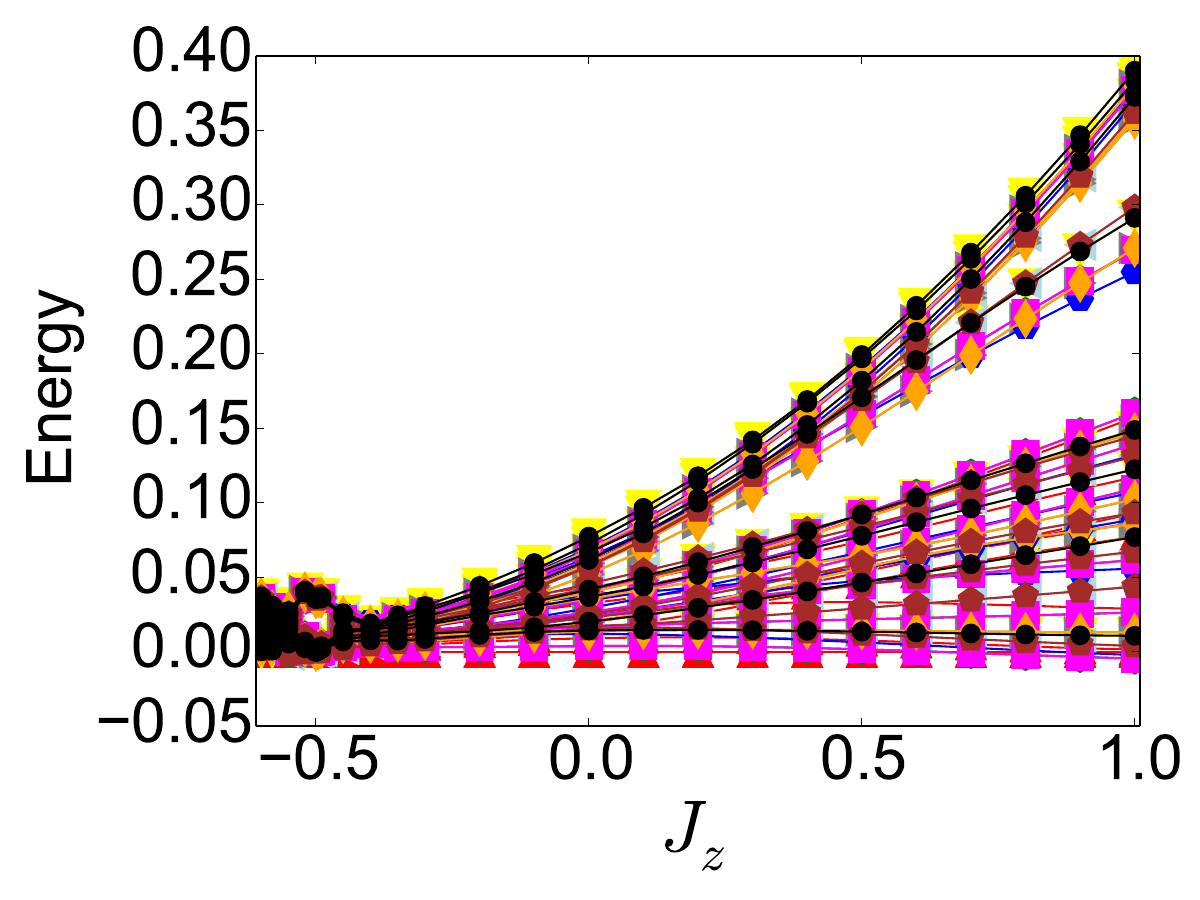}
\includegraphics[width=0.495\linewidth]{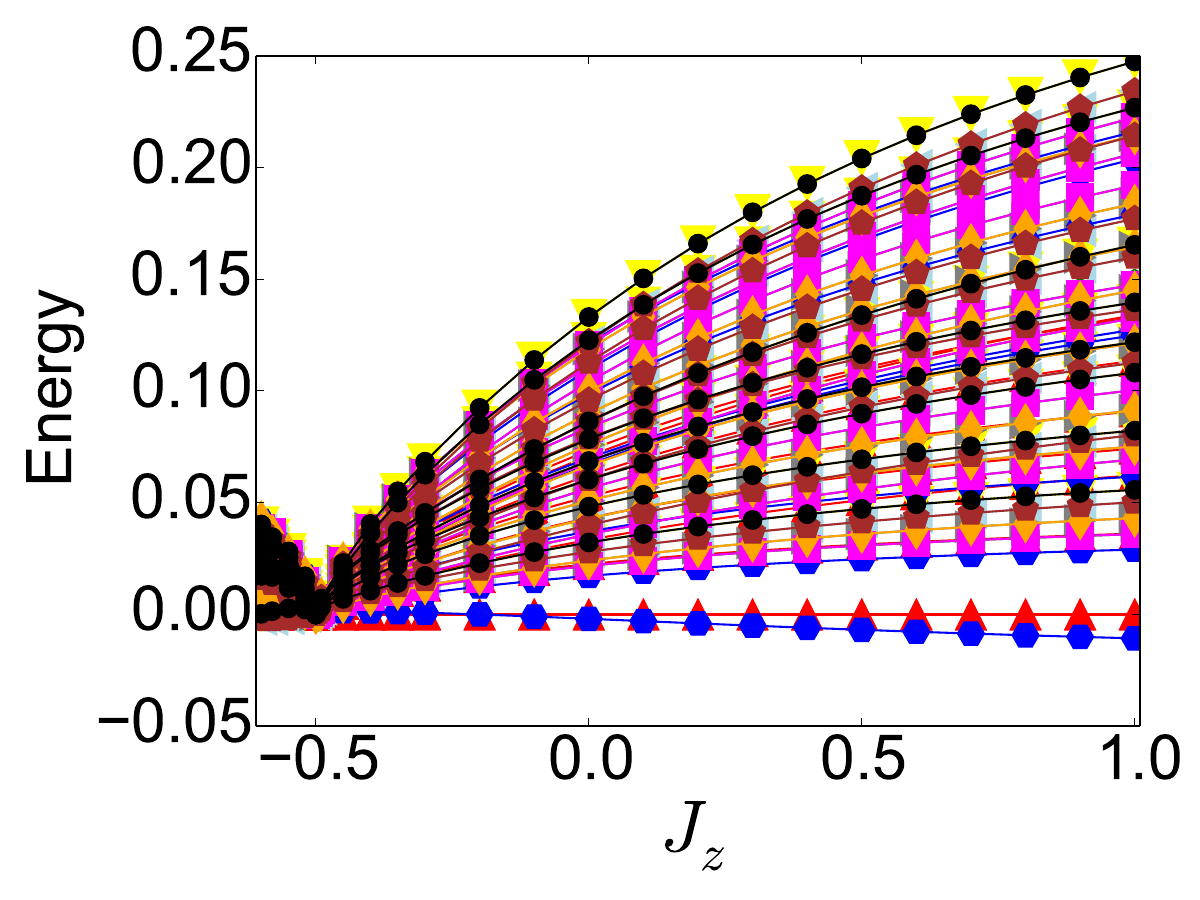}
\includegraphics[width=0.495\linewidth]{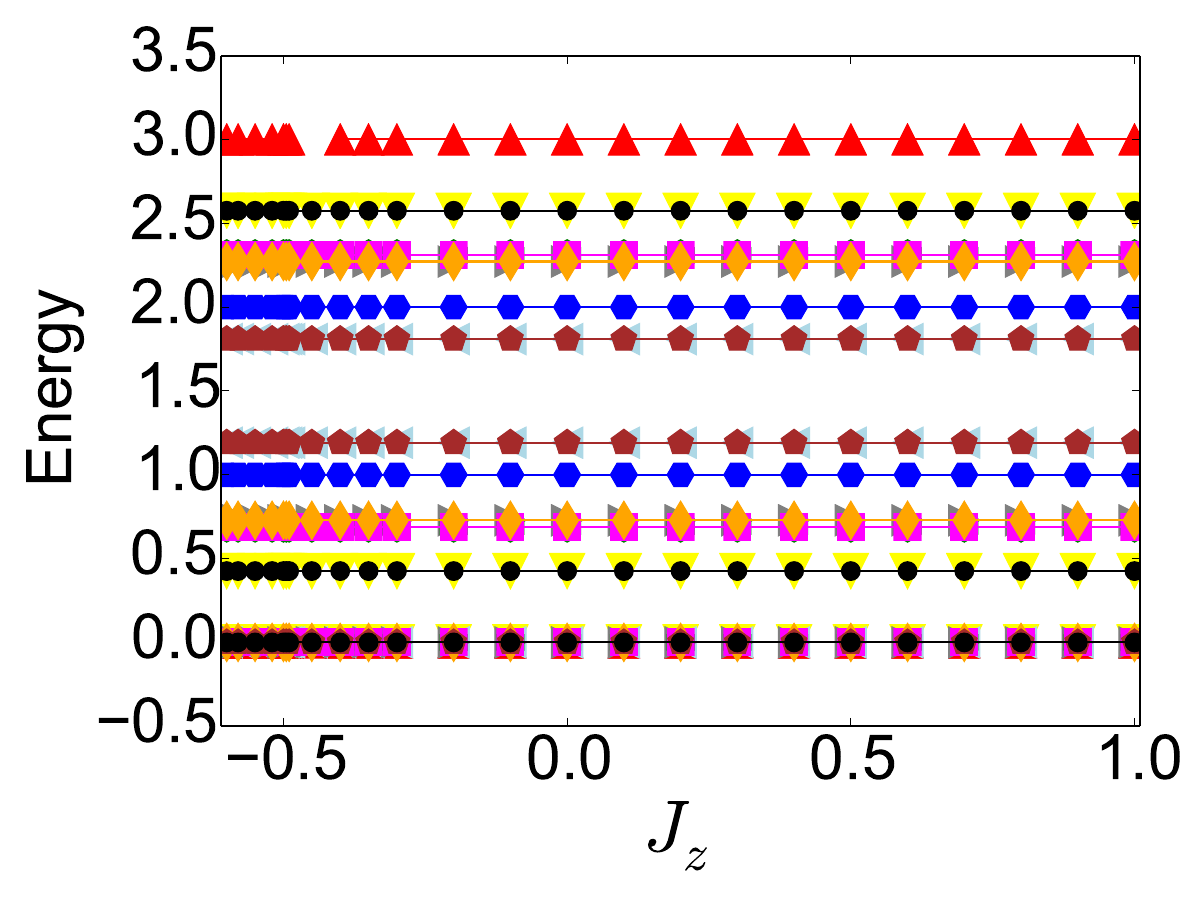}
\caption{(Color online) Energy spectra (showing the 8 lowest energies 
in every momentum sector with respect to the lowest energy state in $K=(0,0)$) 
versus $J_z$ for a 30 site kagome cluster with periodic boundary conditions. 
The panels correspond to various $S_z$ sectors, (top left) $S_z=0$, (top right) $S_z=5$, (bottom left) $S_z=10$,
(bottom right) $S_z=14$. A quantum degeneracy is seen at $J_z=-1/2$. The case of $S_z=14$
corresponds to one spin down in a sea of up spins and maps to 
the non-interacting solution, hence the spectrum does not change with $J_z.$
}
\label{fig:energies} 
\end{figure}	

We first report the existence of a new macroscopic \textit{quantum} degenerate point on kagome and hyper-kagome lattices 
in the spin-1/2 $XXZ$ Hamiltonian~\cite{yamamoto2014quantum,Sellmann,Chernyshev,Richter,Kumar2014,KumarChanglani2016}, 
\begin{equation}
	H_{XXZ}[J_z] =  \sum_{\langle i,j \rangle} S^{x}_{i} S^{x}_{j} + S^{y}_{i} S^{y}_{j} + 
	    J_{z}  \sum_{\langle i,j \rangle} S^{z}_{i} S^{z}_{j} 
\label{eq:XXZ}
\end{equation} 
at $H_\textrm{XXZ}[-1/2]$ (notated as $H_{XXZ0}$~\cite{Jaubert2016}). $S_i$ are spin-1/2 operators 
on site $i$, $\langle i,j \rangle$ refer to nearest neighbor pairs and $J_z$ is the Ising coupling. 
The degeneracy exists in all $S_z$ sectors and all finite system sizes. For kagome, we explicitly demonstrate 
this in Fig.~\ref{fig:energies} where we perform an exact diagonalization (ED) on the $N=30$ site 
kagome cluster in different $S_z$ sectors. As we approach $J_z=-1/2$ many eigenstates collapse to the same ground state eigenvalue.  

\begin{figure}
\centering
\includegraphics[width=0.97\linewidth]{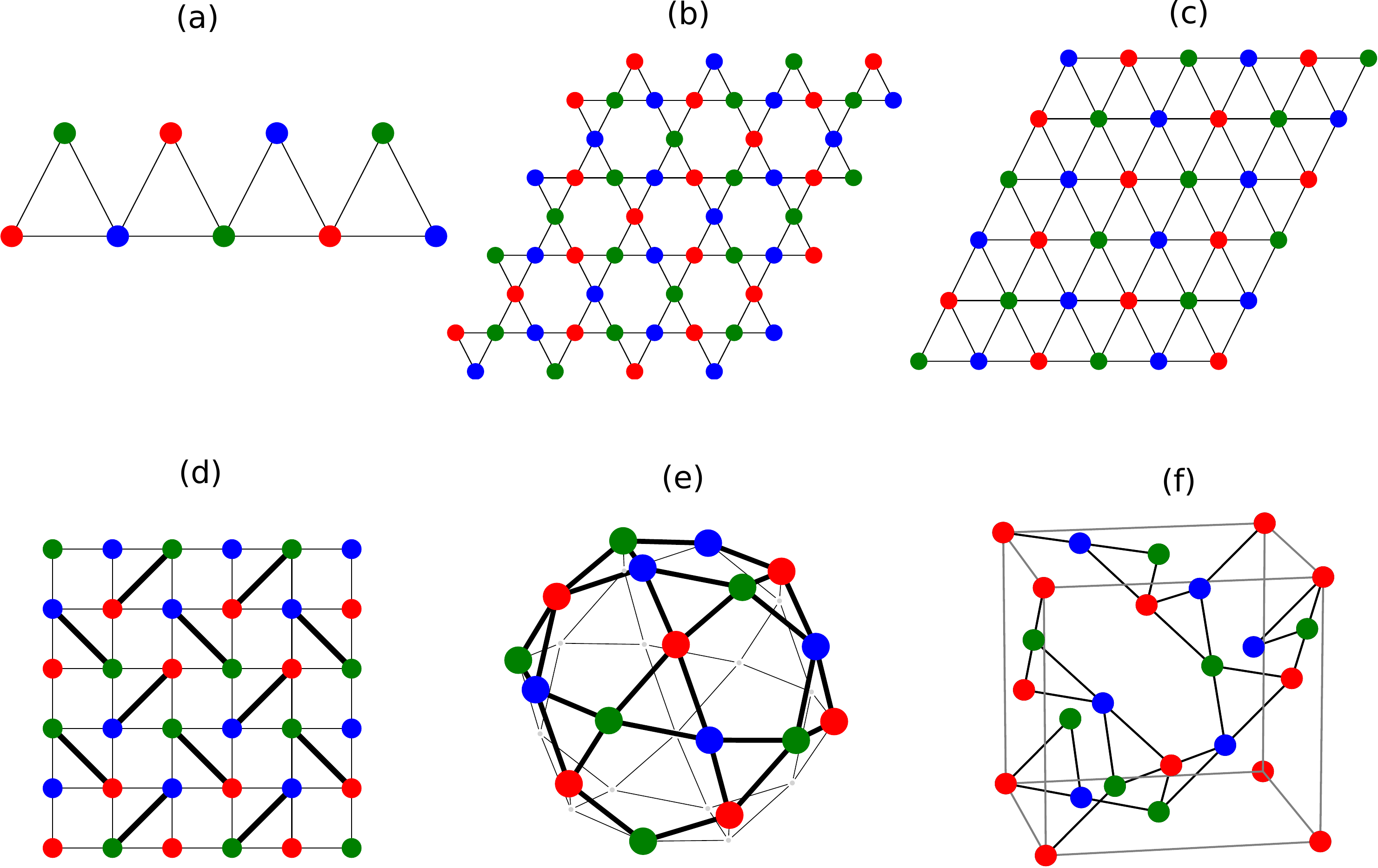}
\caption{(Color online): Representative three-coloring solutions on various lattices 
with triangular motifs. (a) Saw-tooth (b) Kagome
(c) Triangular (d) Shastry-Sutherland~\cite{Shastry_Sutherland} 
(with $J_2=2J_1$, note, the bold diagonal lines are associated with two triangles whereas other edges are part of 
only one triangle) (e) Icosidodecahedron (f) Hyper-kagome lattice}
\label{fig:lattices} 
\end{figure}

We solve analytically for much of the exponential manifold, 
and our solutions apply to any lattice of triangular motifs with the Hamiltonian of the form, 
\begin{equation} 
H = \sum_{\Delta} H_{XXZ0}(\Delta) \label{eq:triangleH}
\end{equation}
where $H_{XXZ0}(\Delta)$ is the $XXZ0$ Hamiltonian on a triangle $\Delta$,
as long as its vertices can be colored by three colors with no two connected vertices 
being assigned the same color. Some three-colorable lattices with representative three-colorings are shown in Fig.~\ref{fig:lattices}. 
Our general result overlaps the $XXZ0$ point on the triangular lattice of Ref.~\cite{Momoi} 
and a different analytically solvable Hamiltonian on the zig-zag ladder of Ref.~\cite{Batista}. 

Finally, we show how the $XXZ0$ point on the kagome lattice is embedded in the wider 
phase diagram demonstrating its relation to the previously discovered spin-liquid at the 
Heisenberg point~\cite{White_Kagome,Depenbrock_Kagome,Jiang_Balents} 
as well as nearby magnetically ordered phases; our results suggest an additional intermediate phase transition in the middle of the spin-liquid region.

\prlsec{Exact Ground States at $J_z = -1/2$ }
Any Hamiltonian of the form of Eq.~\eqref{eq:triangleH} has ground states of the form
\begin{equation}
	| C \rangle \equiv  P_{S_z} \Big( \prod_{\text{valid}} \otimes | \gamma_s \rangle \Big)
\label{eqn:GroundState}
\end{equation}
where $\{ |\gamma_s\rangle = |a\rangle,|b \rangle \text{ or } |c\rangle \}$, denoted as "colors" on site $s$ are defined as, 
$|a \rangle \equiv \frac{1}{\sqrt{2}} \Big(|\uparrow \rangle + |\downarrow \rangle \Big) $,
$|b \rangle \equiv \frac{1}{\sqrt{2}} \Big(|\uparrow \rangle + \omega |\downarrow \rangle \Big)$,
$|c \rangle \equiv \frac{1}{\sqrt{2}} \Big(|\uparrow \rangle + \omega^2 |\downarrow \rangle \Big)$,
where $\omega=e^{i2\pi/3}$. Taking the quantization axis to be the $z$-axis, 
the colors correspond to spin directions in the $XY$ plane that are at $120$ degrees relative to one another. 
Valid colorings satisfy the three-coloring condition. $P_{S_z}$ projects into a particular total $S_z$ sector.   

For $J_z= -1/2$ and a single triangle, 
six states; the fully polarized state $|\uparrow \uparrow \uparrow \rangle$ and 
the chiral states $|\uparrow\downarrow\downarrow \rangle + \omega | \downarrow \uparrow \downarrow \rangle + {\omega}^{2} | \downarrow \downarrow \uparrow \rangle$ and $|\uparrow\downarrow\downarrow \rangle + {\omega}^{2} | \downarrow \uparrow \downarrow \rangle + {\omega} | \downarrow \downarrow \uparrow \rangle$ and all their Kramers pairs; are exactly degenerate. 
Thus Eq.~\eqref{eq:triangleH} is recast as,
\begin{equation}
	H = \sum_{\Delta} H_{\Delta} = \frac{3}{2} \sum_{\Delta} P_{\Delta} - \frac{3}{8} N_{\Delta}
\label{eq:projector_ham}
\end{equation}
where $N_{\Delta}$ is the number of triangles and 
$P_{\Delta}$ is a projector on the triangle $P_{\Delta} \equiv |+ \rangle \langle +| + | - \rangle \langle -|$ 
and $|+\rangle$ and $|-\rangle$ are Kramers pairs of non-chiral one-magnon states on the triangle, $|+ \rangle \equiv  \frac{1}{\sqrt{3}} \Big( |\uparrow \uparrow \downarrow \rangle + | \uparrow \downarrow \uparrow \rangle + | \downarrow \uparrow \uparrow \rangle \Big) $ and $|- \rangle \equiv \frac{1}{\sqrt{3}} \Big( |\downarrow \downarrow \uparrow \rangle + | \downarrow \uparrow \downarrow \rangle + | \uparrow \downarrow \downarrow \rangle \Big)$
This rewriting can be carried out on any lattice of triangles; if a bond is used by multiple triangles this 
constrains the coupling constant between these bonds. 

The $XXZ0$ Hamiltonian is thus a sum of positive semi-definite non-commuting projectors. 
Any wavefunction that simultaneously zeroes out each projector consistently 
is guaranteed to be a ground state. Such "frustration-free" Hamiltonians include 
Majumdar-Ghosh~\cite{Majumdar_Ghosh} (generalized by Klein~\cite{Klein}) 
and Affleck-Kennedy-Lieb-Tasaki~\cite{AKLT,Wen_square,kitaev,Oleg} Hamiltonians. 
Zeroing out a projector requires that only components exactly orthogonal to states $|+\rangle$ and $|-\rangle$ 
enter the full many body wavefunction; this is indeed achieved by the product state 
$|\psi \rangle \equiv  \prod_{\textrm{valid}} \otimes |\gamma_s \rangle$. 
We also note that such "three-coloring states" have a long history and have been explored in several contexts~\cite{Harris3color,Henley3color,
Huse_Rutenberg,Chalker3color,Sachdev92, Cepas, Jaubert2016,Castelnovo_three_coloring}.

The product state $|\psi \rangle$ does not conserve total $S_z$ but the $XXZ$ Hamiltonian \textit{does conserve} it. 
Therefore, projecting each three-coloring solution to each $S_z$ sector is also a ground state leading to 
Eq.~\eqref{eqn:GroundState}. Note that three-colorings which differ simply by relabeling colors are identical up to a global phase (see Supplement).

\begin{table*}
\begin{center}
\begin{tabular}{|c|c|c|c|c|c|c|c|c|c|}
\hline
Lattice     &  $\;$Method$\;$     & $\;\;n_b = 1\;\;$        & $\;\;n_b = 2\;\;$          &  $\;\;n_b = 3\;\;$         & $\;\;n_b = 4\;\;$  &      $\;\;n_b = 5\;\;$      &  $\;\;n_b = 6\;\;$      & $n_b=\lfloor N/2 \rfloor$ & \# 3-colorings  \tabularnewline
\hline
\hline
sawtooth obc           & ED      & 6           & 16           & 26          & 31     & 32        & 32         & 32   & 32   \tabularnewline
$5$ triangles          & $R(S)$  & 6           & 16           & 26          & 31     & 32        & 32         & 32   &      \tabularnewline
\hline
$3\times3$ kagome obc  & ED      & 15          & 102          & 414         & 1117    &           &           &      & 3808 \tabularnewline
(33 sites)             & $R(S)$  & 15          & 102          & 414         & 1117    &  2136     &  3078     & 3808 &      \tabularnewline
\hline
$3\times3$ kagome pbc  & ED      & 10          & 38           & 60          & 41      & 40       & 40         &  40  & 40          \tabularnewline
                       & $R(S)$  & 10          & 34           & 40          & 40      & 40       & 40         &  40  &      \tabularnewline
\hline
$4\times3$ kagome pbc & ED       & 13         & 68           & 169         & 172     & 137       & 136        &      & 136  \tabularnewline
                      & $R(S)$   & 13         & 68           & 134         & 136     & 136       & 136        &  136 &      \tabularnewline
\hline
\hline
\end{tabular}
\caption{
Number of ground states in different $S_z$ sectors (mapped to hard-core boson number $n_b$)
on several lattices (of size $N$) with triangular motifs at $J_z=-1/2$, $J_2 = 0$. 
$R(S)$ is the rank of the overlap matrix indicating the number of linearly independent 3-coloring modes 
and ED refers to the exact number of ground states. 
The kagome cluster with open boundary conditions (obc) has completed triangles, 
resembling the periodic counterpart (pbc) in appearance. 
}
\label{tab:colors}
\end{center}
\end{table*}

\prlsec{Macroscopic Degeneracy and additional ground states}
While there are only two ways of three-coloring the triangular lattice, 
there are an exponential number of ways of doing so on the kagome (scaling as $1.208^N$ \cite{Baxter1970}) and hyper-kagome lattices. 
The precise number of ground states varies from sector to sector because of the loss of linear 
independence of the unprojected solutions under projection. 
For typical $S_z$ of interest, particularly $S_z=0$, 
there are still an 
exponential number of linearly independent solutions. 
This counting is made precise by forming the overlap matrix $S_{C,C'} \equiv \langle C | C' \rangle$ 
and evaluating its rank $\equiv R(S)$ numerically; our results have been shown in Table~\ref{tab:colors} and the Supplement.
The case of one down spin in a sea of up spins which maps to the non-interacting 
problem with a flat-band with a quadratic band touching~\cite{Bergman_Balents} is also correctly captured. 

On several representative clusters with open boundary conditions (but always with completed triangles), we never find solutions outside the coloring manifold 
which suggests (but does not prove) the possibility that coloring solutions describe all degeneracies on open lattices. 
However, for kagome on tori we find, for low fillings, degenerate solutions not spanned by colorings.

\begin{figure}
\centering
\includegraphics[width=\linewidth]{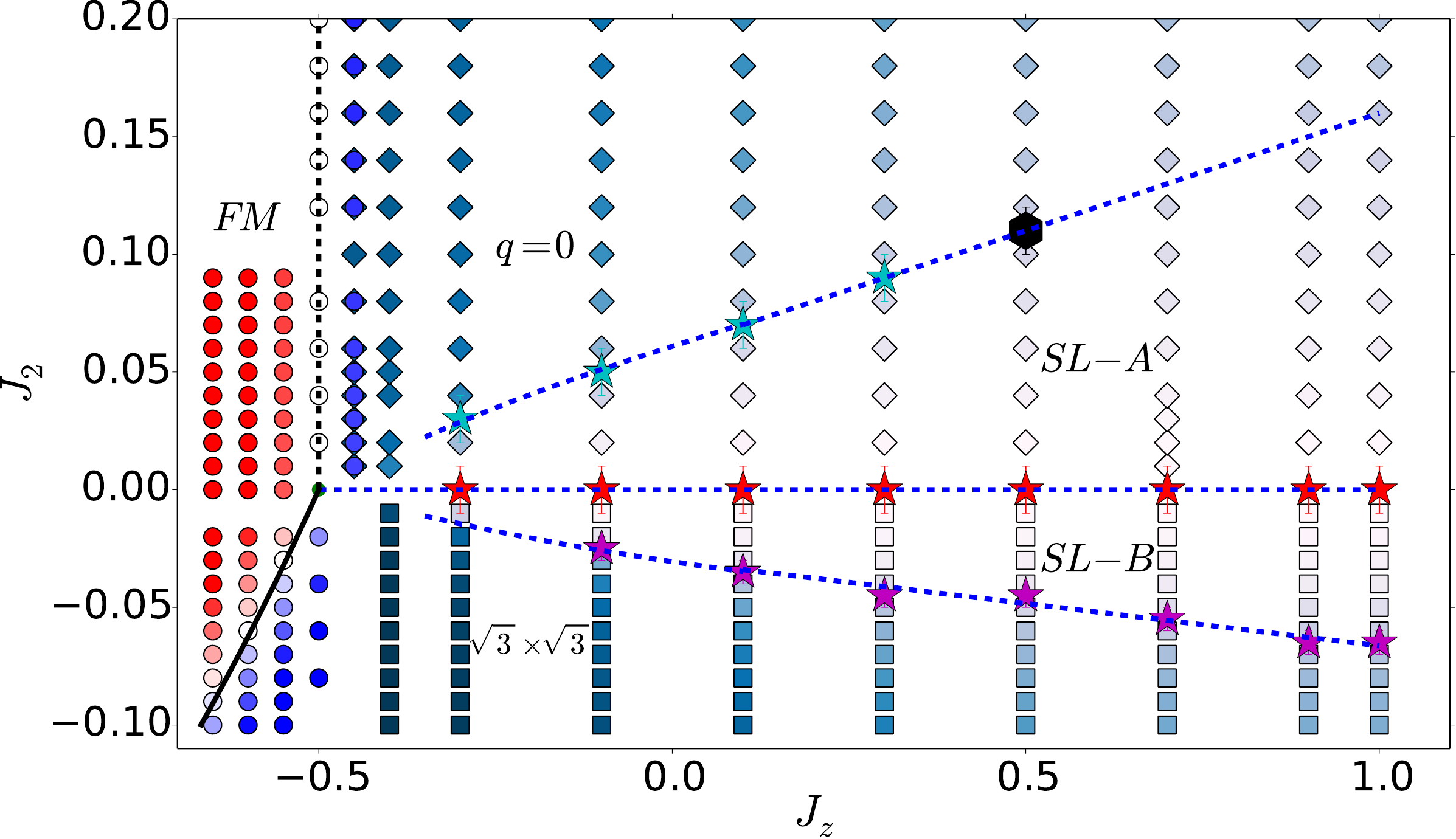}
\caption{(Color online): The phase diagram in the $J_z - J_2$ plane on the 36d lattice showing five phases - 
the ferromagnet (FM), the magnetically ordered phases ($q=0$ and $\sqrt{3}\times\sqrt{3}$), and the spin liquids (SL-A and SL-B). 
Circles correspond to the energy difference $E(S_z=0)_{N=36}-E_{TDL}(S_z=N/2)$ between the $S_z=0$ sector and 
fully polarized state ranging from deep blue (negative) to deep red (positive).  
The diamonds are colored based on the structure factor at the $M$ point ($S(M)$) and squares are colored based on the structure factor 
at the $K$ point ($S(K)$).  The darkest color corresponds to the largest structure factor on the graph. 
Star symbols correspond to location of fidelity dips 
and the error-bars indicate the uncertainty in the location of the phase boundaries (when scanned in the $J_2$ direction) 
and correspond to the grid-spacing used for the computation of the fidelity.
The black hexagon (at $J_z\approx0.5$, $J_2 \approx 0.10 $) is a kink in the second derivative of the fidelity; 
beyond the corresponding $J_z$ the fidelity dip is not noticeable and the phase boundary is just an extrapolation.  
Phase boundaries are marked with dotted lines, which are guides to the eye. 
The solid line is where the semiclassical energy difference between the FM 
and the unprojected $\sqrt{3} \times \sqrt{3}$ state goes to zero. 
}
\label{fig:kagome} 
\end{figure}

\prlsec{Connection to the wider Kagome phase diagram} 
We now show how the $XXZ0$ point is embedded in the larger kagome phase diagram. 
We focus on $S_z=0$ and the fully symmetric sector of $K=(0,0)$ sector (see Supplement), 
and study an extended Hamiltonian involving nearest neighbor (nn) and next-nearest neighbor (nnn) terms,
\begin{equation}
	H[J_z,J_2] = H^\textrm{nn}_{XXZ}[J_z] + J_2 H^\textrm{nnn}_{XXZ}[J_z]
\end{equation}
where $H^\textrm{nnn}_{XXZ}[J_z] = \Big(\sum_{\langle \langle i,j \rangle \rangle} S^{x}_i S^{x}_j +S^{y}_i S^{y}_j + J_z S^{z}_i S^{z}_j \Big)$; 
$\langle\langle i,j \rangle\rangle$ referring to nnn pairs. We use a combination of analytical arguments and 
ED on the 36d cluster~\cite{Elser93,SudanED} on a grid of points in the $(J_z,J_2)$ space. 
As Fig.~\ref{fig:kagome} shows, we find five phases near $XXZ0$: 
a ferromagnetic phase, a $q=0$ phase, a $\sqrt{3} \times \sqrt{3}$ phase and (potentially) two spin-liquids. 
We give numerical evidence that all these phases, other than the ferromagnet, connect from near (or touching) XXZ0 to the Heisenberg point. 


At $J_z=-1/2$ and $J_2>0$, (notated AF-line) 
all triangles in the Hamiltonian are of the $XXZ0$ form and remain 
consistently three-colorable. Three-coloring both nn and nnn triangles constrains the allowed colorings 
leaving only two colorings in the well known $q=0$ pattern.
This phase survives for $J_z>-1/2$, at small $J_2$, and is primarily identified by peaks at the $M$ point 
(Fig. S2 of the Supplement) in the spin-structure factor $S(\vec{q}) \equiv \frac{1}{N}\sum_{i,j} \langle S_{i} \cdot S_{j} \rangle e^{i\vec{q} \cdot (\vec{r_{i}}-\vec{r_{j}})} $ where 
$\vec{r}_{i}$ refers to the real space coordinates of the $i^{th}$ lattice site, $N$ is the total number of sites and $ \langle S_{i} \cdot S_{j} \rangle $ 
is the spin-spin correlation function. On the other hand, it can be 
rigorously shown the minimum energy state upon 
perturbing the AF-line to $J_z<-1/2$ is the fully polarized ferromagnetic state. 

At $J_z=-1/2$ and $J_2<0$, we find evidence for the $\sqrt{3} \times \sqrt{3}$ phase. 
While we can not solve for the exact ground state, the state which colors nnn triangles the same color 
(i.e. the $\sqrt{3} \times \sqrt{3}$  phase)  minimizes the nnn energy  within the three-coloring manifold.  
We numerically verify this phase by looking at $S(K)$, finding it survives for $J_z$ near and on both sides of $-1/2$.

\begin{figure}
\centering
\includegraphics[width=\linewidth]{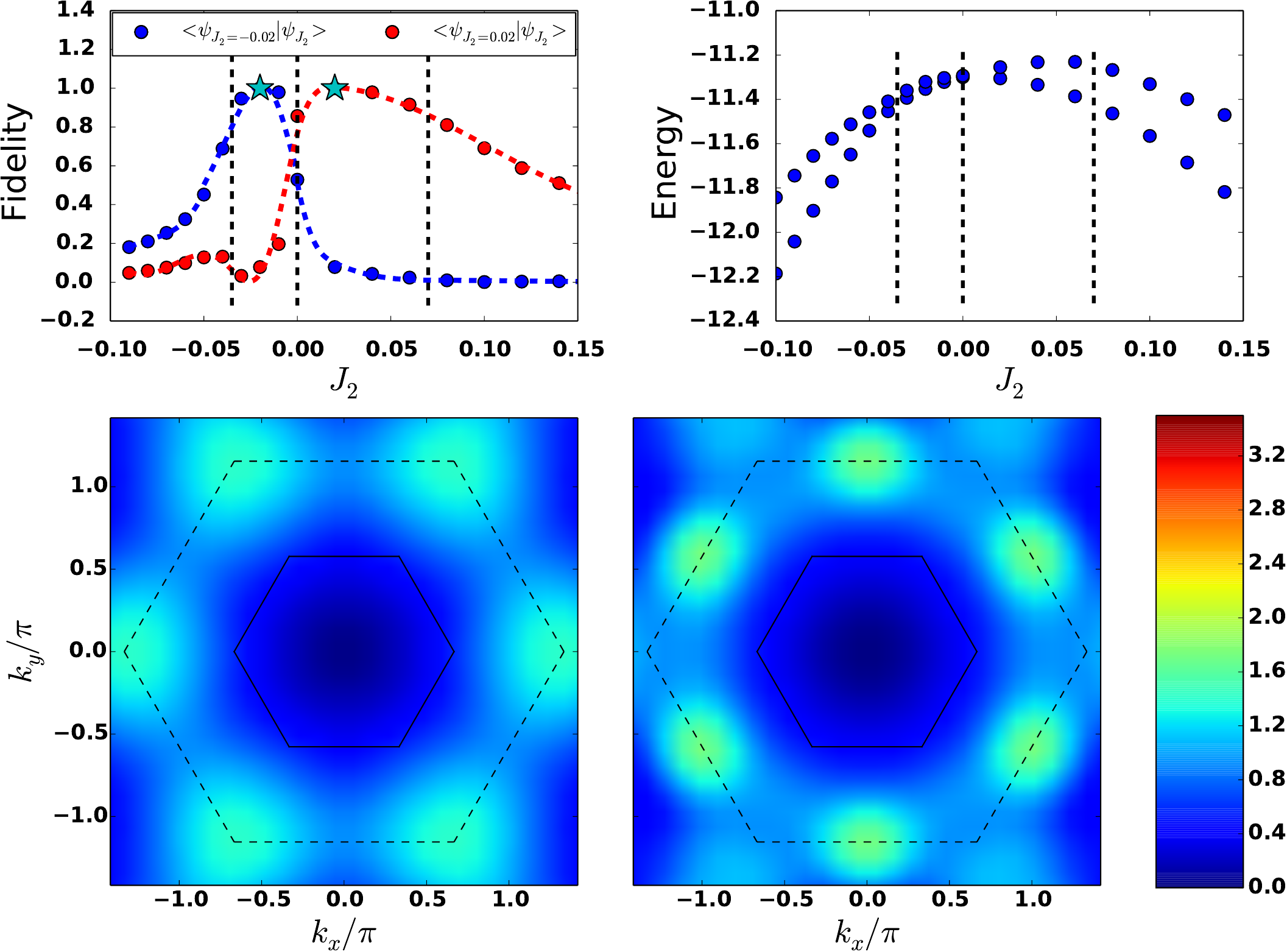}
\caption{ (Color online) All data is at $J_z=0.1$ for the 36d lattice.  Top Left: Overlap of the ground state at $J_2$ 
with respect to reference ground state wavefunctions at $J_2=-0.02$ (blue) and $J_2=0.02$ (red). 
Dashed lines represent transitions as measured by fidelity. Top Right: Energy of 
the two lowest states in the symmetric representation of the $K=(0,0)$ sector. There are additional state(s) 
between these two states in other quantum-number sectors. 
Bottom: The static spin structure factor $S(\vec{q})$ of the ground state
for $J_2=-0.02$ (left) and $J_2=0.02$ (right). 
The solid and the dotted lines show the first and the extended Brillouin zones 
respectively. The high symmetry points of the latter correspond to 
$K$ (corners of the hexagon) and $M$ (midpoints of edges) points. On going from $J_2<0$ to $J_2>0$, 
the intensity is transferred from $K$ to $M$ points. }
\label{fig:dips} 
\end{figure}	

By tracing paths through parameter space with large values of $S(\vec{q})$ at the $K$ and $M$ points, 
we find that both the $q=0$ phase 
and $\sqrt{3} \times \sqrt{3}$ phases near the $XXZ0$ point extend to the Heisenberg point at non-zero $J_2$. 
To locate the boundaries of these phases, 
we perform sweeps through $J_2$ at fixed $J_z$ and identify dips in the 
wavefunction fidelity defined to be
\begin{equation}
    f(J_z, J_2) \equiv \Big| \langle \psi (J_z, J_2 - \Delta J_2/2 )|\psi (J_z,J_2 + \Delta J_2/2) \rangle \Big| \\
\end{equation}
where $\psi(J_z,J_2)$ is the ground state wavefunction, $\Delta J_2$ is the step size in the $J_2$ direction. For both magnetically ordered phases, the location of these dips form 
lines emanating from (or close to) the $XXZ0$ point that extrapolate to the Heisenberg point ($J_z=1$) 
to values $J_2 \approx 0.16$ for $q=0$ and $J_2 \approx -0.06$ for $\sqrt{3} \times \sqrt{3}$. These values 
are within the bounds previously found by a DMRG study~\cite{kolley2015phase}, but disagree with a variational study by Ref.~\cite{Iqbal_Becca_NJPhys} which finds 
instead a valence bond crystal. In the intermediate phase(s), we see a decrease in the magnitude of the structure factor peaks consistent with a change in phase to a spin-liquid.

Near $XXZ0$ we do not detect fidelity dips and see larger structure factors that extend much 
closer to the line $J_2=0.$  This leaves two plausible scenarios: (1) the spin-liquid(s) terminate at $J_z>-1/2$ 
for all $J_2$ or (2) the phase boundaries extend to $XXZ0$ but finite size-effects near it become large making it difficult 
to resolve the transition.

We find an additional fidelity dip at $J_2 \approx 0$ and $J_z>-1/2$ in the region where other studies~\cite{kolley2015phase} identify a 
single spin-liquid phase.  This interesting finding indicates the existence of an additional transition in this region. Our analysis in this
work is largely ambivalent about the nature of these two phases but earlier evidence for a spin-liquid phase at $J_z=1$ and 
both $J_2>0,J_2<0$~\cite{kolley2015phase,Normand_Xiang} suggests a possible transition between two spin-liquids.  Interestingly, a recent IPEPS study~\cite{jiang2016competing} found nearly degenerate variational degenerate energies for the $Q_1=Q_2$ and $Q_1=-Q_2$~\cite{Sachdev92} Z$_2$-spin liquids which they interpret as evidence for a parent U(1) DSL; given our results, another reasonable interpretation is that there is a transition between these two states.

To further understand the nature of the fidelity dips, 
we consider the ground state and excited state in the same quantum number sector as a function of $J_2$ at $J_z=0.1$ 
(Fig.~\ref{fig:dips}, top right); the true first excited-state is in another sector. 
We see a (formally avoided) "level-crossing" indicated by a shrinking gap between these states around $J_2 \approx 0$. 
This crossing causes the fidelity dip and leads to the overlap of the wavefunction on both sides of $J_2 \approx 0$ being small 
with respect to a reference point on the other side (see Fig.~\ref{fig:dips}, top left). 
In addition, the structure factors of the two ground states at positive and negative $J_2$, despite not 
having large peaks, are qualitatively distinct (see Fig.~\ref{fig:dips}, bottom).

\prlsec{Conclusion}
In summary, we have (1) shown that $H_\textrm{XXZ0}$ is macroscopically quantum degenerate on the kagome and 
hyperkagome lattices, (2) shown that all projected three-coloring states are exact ground states of $H_\textrm{XXZ0}$ on any three-colorable lattice of 
triangular motifs explaining this macroscopic degeneracy, (3) shown that multiple phases  in the $J_2-J_z$ phase diagram, including the spin-liquid(s) in the Heisenberg regime, 
are proximate to the $XXZ0$ point, and (4) given evidence for a transition between two phases at $J_2=0$  for $-0.5<J_z<1$.  
Our findings suggest that the $XXZ0$ point controls the physics of the Heisenberg and $XY$ points~\cite{YCHeXXZ,LauchliMoessner} on the kagome and the existence of a transition near the Heisenberg point might help resolve conflicting numerical evidence for gapless and gapped states respectively.  
While our focus here has been on the uniform kagome lattice, the exponential degeneracy also applies in the case where the coupling constant in each triangle is disordered (or staggered) as well as to finite clusters of triangles such as the icosidodecahedron; in fact, the latter explains the nearly degenerate manifold on this cluster in the $XY$ regime~\cite{icosi}. 

The central coloring ideas extend to other frustrated lattices with four (or higher) site motifs~\cite{KondevHenley,Khemani2012,WanGingras2016}. 
For example, define a Hamiltonian which annihilates four-coloring states made of one
$a \equiv |\uparrow\rangle + |\downarrow\rangle$, $b \equiv |\uparrow\rangle+ i |\downarrow\rangle$,
$c \equiv |\uparrow\rangle - |\downarrow\rangle$ and $d \equiv |\uparrow\rangle -i |\downarrow\rangle$
on each square of a square lattice or tetrahedron of the pyrochlore lattice. Up to a constant, this is 
$H = 2 H_{XXZ}[-1/4] + \sum_{i<j, k<l, \text{diff}}  S_i^{+}S^{+}_j  S_k^{-}S^{-}_l  - 2\; S^{z}_{1} S^{z}_{2} S^{z}_{3} S^{z}_{4}$
where ``diff'' indicates $i, j, k, l$ are distinct (see Supplement for the derivation that used the DiracQ package~\cite{DiracQ}). Notice that 
on the square this forces the nnn $J_2$ coupling to be half the nn $J_1$ coupling; interestingly $J_2/J_1=1/2$ has been proposed to be a SL state on 
the square for Heisenberg and XY models~\cite{Duan}.We believe that the macroscopic degeneracy of this Hamiltonian on the square and pyrochlore 
lattices will be a source of multiple phases on these lattices~\cite{Normand,Hermelepyrochlore}.

Finally, we note that three-coloring states can be used to construct accurate many-body wavefunctions~\cite{Huse_Elser,Changlani_CPS, NeuscammanCPS, Tay_Motrunich}. 
Typically Jastrow factors have been introduced only on top of a single coloring; our present investigation 
suggests that a linear combination of colorings may provide accurate results in the vicinity of 
the $XXZ0$ point. 

\prlsec{Acknowledgement}
We thank V. Elser, S. Shastry, O. Tchernyshyov, V. Chua, L.D.C Jaubert, S. Sachdev, R. Flint, P. Nikolic, Y. Wan 
and O. Benton for discussions and H. Wang for collaboration on related work. We also thank 
T. Momoi for bringing to our attention Ref.~\cite{Momoi} after this work was posted. HJC, DK and BKC were supported by SciDAC grant 
DE-FG02-12ER46875 and KK and EF by NSF grant numbers DMR 1408713 and 1725401. 
HJC also acknowledges funding from the U.S. Department of Energy, Office of Basic Energy Sciences, Division of Materials Sciences and Engineering under Award DE-FG02-08ER46544 for his work at the Institute for Quantum Matter (IQM). This research is part of the Blue Waters sustained petascale computing project, 
which is supported by the National Science Foundation (award numbers OCI-0725070 and ACI-1238993) and the State of Illinois.

\bibliography{refs}

\pagebreak
\widetext
\textbf{\large Supplemental Material for "The mother of all states of the kagome quantum antiferromagnet"} 
\setcounter{equation}{0}
\setcounter{figure}{0}
\setcounter{table}{0}
\setcounter{page}{1}
\makeatletter
\renewcommand{\theequation}{S\arabic{equation}}
\renewcommand{\thefigure}{S\arabic{figure}}
\renewcommand{\bibnumfmt}[1]{[S#1]}
\renewcommand{\citenumfont}[1]{S#1}

\section{Efficient Overlap and Hamiltonian Matrix elements in the 3-coloring basis}

In the main text, we mentioned the efficient evaluation of the number of linearly independent 
colorings when projected to definite total $S_z$ (whose value we denote as ${S_z}^{*}$). This number was 
obtained by diagonalizing the overlap matrix and determining its rank. 
Here we present expressions for the overlap and Hamiltonian matrices in the $S_z$ (or number, 
in the hard-core boson language) projected-coloring basis which correspond to $S_{CC'} \equiv \langle C|C' \rangle$ and 
$H_{CC'} \equiv \langle C | H |C'\rangle$ 
respectively. A projected coloring $|C\rangle$ is given by the expression,
\begin{equation}
	| C \rangle \equiv  P_{S_z} \Big( \prod_{i} \otimes | c_i \rangle \Big)
\label{eqn:pcolor}
\end{equation}
where $|c_i \rangle$ is the color on site $i$ and can be $|a\rangle$, $|b\rangle$ or $|c\rangle$, as defined in the main text.

Matrix elements involving projected-colorings are calculated by introducing a complete set of orthonormal 
states, which for the present purpose is chosen to be the Ising basis, compactly written as, 
\begin{equation}
	I \equiv \{ s_1, s_2, s_3, ....s_N\}
\end{equation}
where $s_i$ are Ising variables with value $\pm \frac{1}{2}$ on site $i$, and $N$ is the total number of sites. 
Introducing the identity operator we have, 
\begin{equation}
	\langle C|C' \rangle  = \sum_{I \text{in sector}}\langle C | I \rangle \langle I|C' \rangle
\end{equation}
Naively, this summation may be evaluated only by enumerating all Ising configurations in a given spin sector ($S_z^{*}$) 
and will thus take an exponentially increasing amount of time to evaluate. However, the Ising sum can be converted to 
one over unconstrained variables $s_1,s_2,...s_N$ and the summation becomes very easy to compute as it 
factorizes into a product of sums. This is achieved by introducing a delta function and then Fourier transforming the expression
as follows,
\begin{subequations}
\begin{eqnarray}
	\langle C|C' \rangle    &=& \sum_{I \text{ unconstrained} }\langle C | I \rangle \langle I|C' \rangle \delta(S_z - S_z^{*}) \\
				&=& \frac{1}{N+1} \sum_{p} \sum_{I}\langle C | I \rangle \langle I|C' \rangle e^{\text{i}p(S_z-S_z^{*})} \\
				&=& \frac{1}{N+1} \sum_{p} \prod_{j} \sum_{s_j} e^{\text{i} p s_j} \langle c_j |s_j \rangle \langle s_j |c_j'\rangle e^{-\text{i}pS_z^{*}}
\end{eqnarray}
\end{subequations}
where the sum over $p$ ranges from $p=0$ to $p=2\pi N/{N+1}$ in multiples of $2\pi/{N+1}$. 
This is because $S_z$ varies from a minimum of $-N/2$ to a maximum of $N/2$. 
Note that we have used $S_z = s_1 + s_2 + s_3...+s_N$ to factorize the product into a product of sums.

Associating integers $0$,$1$,$2$, with the colors $a,b,c$ respectively,
it follows that,
\begin{subequations}
\begin{eqnarray}
	\langle s_j | c_j \rangle &=& \frac{1}{\sqrt{2}} \omega^{\Big( c_j/2 - c_j s_j \Big)} \\
	\langle c_j | s_j \rangle &=& \frac{1}{\sqrt{2}} \omega^{\Big( c_j - 2c_j s_j \Big)}
\end{eqnarray}
\end{subequations}
where $\omega \equiv e^{\text{i} 2\pi /3}$. In order to simplify the expression of the overlap, we define the variables, 
\begin{equation}
	\lambda_j \equiv (2c_j + c_j') \text{(mod 3)} = (c_j' - c_j) \text{(mod 3)}
\end{equation}
and the function, 
\begin{equation}
	f^0(p,\lambda_j) \equiv \frac{1} {2} ( e^{\text{i} p/2} + e^{\text{i} 2\pi\lambda_j/3} e^{-\text{i}p/2} )
\end{equation}
Thus the overlap matrix element reads,
\begin{equation}
	\langle C|C' \rangle  = \frac{1}{N+1} \sum_{p} F^0(p)
\end{equation}
where we have defined,
\begin{equation}
	F^0(p) \equiv \prod_{j} f^0(p,\lambda_j) e^{-\text{i}pS_z^{*}}
\end{equation}
This equation is correct only up to a normalization factor, because the definition of $C$ and $C'$ 
does not guarantee an overall normalization automatically. This normalization is just the combined 
weight on all configurations in the full (unprojected) Hilbert space divided by the combined 
weight on the configurations in the correct $S_z$ sector. 
Including all prefactors into one term we define, 
\begin{equation}
	\mathcal{N} = \frac{1}{N+1} \times \frac{ 2^{N} } {\text{Total Ising configurations in correct sector}}
\end{equation} 
which makes the expression for the overlap, 
\begin{equation}
	\langle C|C' \rangle  = \mathcal{N} \sum_{p} F^0(p)
\end{equation}
A similar delta function trick can be used in the evaluation of the Hamiltonian matrix elements. 
For example, the diagonal element in the $S_z$ basis 
is $S^{z}_m S^{z}_n$  and can be evaluated as, 
\begin{equation}
	\langle C| S^{z}_m S^{z}_n | C' \rangle  = \mathcal{N} \sum_{p} \Big( \frac{f^z(p,\lambda_m) f^z(p,\lambda_n)}{f^0(p,\lambda_m) f^0(p,\lambda_n)} \Big) F^0(p)
\end{equation}
where
\begin{equation}
	f^z(p,\lambda_j) \equiv \frac{1} {4} ( e^{\text{i} p/2} - e^{\text{i} 2\pi\lambda_j/3} e^{-\text{i}p/2} )
\end{equation}

The off diagonal element is also straightforward and is found to be, 
\begin{equation}
	\langle C| S^{+}_m S^{-}_n | C' \rangle  = \mathcal{N} \sum_{p} \Big( \frac{f^{+}(p,c'_m) f^{-}(p,c_n)}{f^0(p,\lambda_m) f^0(p,\lambda_n)} \Big) F^0(p)
\end{equation}
where
\begin{subequations}
\begin{eqnarray}
	f^{+}(p,c_j) \equiv \frac{1} {2} e^{\text{i} 2\pi  c_j/3} e^{\text{i}p/2} \\ 
	f^{-}(p,c_j) \equiv \frac{1} {2} e^{\text{i} 4 \pi c_j/3} e^{-\text{i}p/2} 
\end{eqnarray}
\end{subequations}
These last two expressions do not depend on $\lambda_j$ but rather the value of the color in the ket or bra.

\section{Counting the number of three-colorings}

In Table I of the main paper, we showed the number of valid 3-colorings (i.e. colorings which 
satisfied the constraint of one distinct color per triangular motif) for several lattices. 
The counting was automated employing a simple divide and conquer algorithm. The lattice was divided 
into $P$ pieces, and for each piece the number of valid 3-colorings was checked by brute 
force enumeration of configurations. Then the 3-coloring consistency condition between pieces was 
checked and the combinations were retained or eliminated accordingly. In practice, for the small lattices considered here, 
$P=1$ to $P=6$ sufficed, but for larger lattices larger $P$ is possibly 
needed for efficient counting. 

In order to not over-count colorings, it is important to fix the color 
of one (reference) site to $a$ in all valid colorings. This is 
because the coloring $C'$, obtained by exchanging the colors (consistently for all sites) 
of a coloring $C$, is not linearly independent of it. This can be seen by redefining,
\begin{equation}
	|\downarrow \rangle' \equiv \omega |\downarrow \rangle
\end{equation} 
which is equivalent to the transformation (from old to new variables)
\begin{subequations}
\begin{eqnarray}
a \rightarrow c \\
b \rightarrow a \\ 
c \rightarrow b
\end{eqnarray} 
\end{subequations}
Under this transformation each spin configuration (and hence the overall wavefunction) 
is simply rescaled by a constant factor of $\omega^{N_{\downarrow}}$ where $N_{\downarrow}$ 
is the number of down spins. (A similar transformation holds for $|\downarrow\rangle' \equiv \omega^2 |\downarrow \rangle$ 
which leads to $a \rightarrow b$, $b \rightarrow c$, $c \rightarrow a$). Thus, 
these colorings are not linearly independent and should not be (double or triple) counted.

In Table~\ref{tab:clusters}, we show several finite clusters (including those shown in the main text) 
where the number of 3-colorings were computed 
and show their correspondence with the number of ground states found from exact diagonalization (ED). The 
number of linearly independent colorings is the rank ($R(S)$) of the overlap matrix ($S_{CC'}$), whose efficient
evaluation was discussed in the previous section.

\begin{table}[htpb]
\begin{center}
\begin{tabular}{|c|c|c|c|c|c|c|c|c|}
\hline
Lattice     & $\;$Method$\;$     & $\;\;n_b = 1\;\;$        & $\;\;n_b = 2\;\;$          &  $\;\;n_b = 3\;\;$         & $\;\;n_b = 4\;\;$  &      $\;\;n_b = 5\;\;$      &  $\;\;n_b = 6\;\;$      & \# 3-colorings  \tabularnewline
\hline
\multicolumn{9}{c} {Finite clusters} \\
\hline
sawtooth obc       & ED      & 6           & 16           & 26            & 31          & 32        & 32        & 32         \tabularnewline
length 5       & $R(S)$    & 6           & 16           & 26            & 31          & 32        & 32        &            \tabularnewline
\hline
Husimi cactus   & ED      & 5           & 11           & 15            & 16          & 16        & 15        & 16         \tabularnewline
generation$=1$           & $R(S)$    & 5           & 11           & 15            & 16          & 16        & 15        &            \tabularnewline
\hline
$21$ site kagome          & ED      & 10          & 44           & 112           & 187         &  231      & 243       & 244        \tabularnewline
            & $R(S)$    & 10          & 44           & 112           & 187         &  231      & 243       &            \tabularnewline
\hline
$3\times2$ kagome obc      & ED      & 11          & 54           & 156           & 299         &  418       & 474       & 488        \tabularnewline
(23 sites)     & $R(S)$    & 11          & 54           & 156           & 299         &  418       & 474       &            \tabularnewline
\hline
$3\times3$ kagome obc      & ED      & 15          & 102          & 414           &  1117       &            &           & 3808        \tabularnewline
(33 sites)     & $R(S)$    & 15          & 102          & 414           &  1117       &  2136      &  3078     &            \tabularnewline
\hline
\multicolumn{9}{c} {Kagome on tori} \\
\hline
$2\times2$  & ED      & 5          & 8            & 8            & 8          & 8         & 8        & 8          \tabularnewline
	    & $R(S)$    & 5          & 8            & 8            & 8          & 8         & 8        &            \tabularnewline
\hline
$3\times2$  & ED      & 7          & 17           & 17           & 16         & 16        & 16       & 16         \tabularnewline
            & $R(S)$    & 7          & 15           & 16           & 16         & 16        & 16       &            \tabularnewline
\hline
$4\times2$  & ED      & 9          & 30           & 42           & 33         &  32       & 32       & 32         \tabularnewline
            & $R(S)$   & 9          & 26           & 31           & 32         &  32       & 32       &             \tabularnewline
\hline
$5\times2$  & ED      & 11         & 47           & 92           & 83         &  65       & 64       & 64         \tabularnewline
            & $R(S)$    & 11         & 42           & 58           & 63         &  64       & 64       &            \tabularnewline
\hline
$3\times3$  & ED      & 10         & 38           & 60           & 41         &  40       & 40       & 40         \tabularnewline
            & $R(S)$    & 10         & 34           & 40           & 40         &  40       & 40       &            \tabularnewline
\hline
$4\times3$  & ED      &	13         & 68           & 169          & 172        &  137      & 136      & 136        \tabularnewline
            & $R(S)$   & 13         & 68           & 134          & 136        &  136      & 136      &            \tabularnewline
\hline
$4\times4$  & ED      &	17	   & 122          & 459          & 875        &  793      & -        & 720         \tabularnewline
            & $R(S)$    &	17	   & 122          & 447          & 683        &  719      & 720      &             \tabularnewline
\hline
\end{tabular}
\caption{Number of ground states on lattices with triangular motifs calculated from the rank of the overlap 
matrix of 3-colorings ($R(S)$) and from exact diagonalization (ED). For all studied clusters and number 
of hard-core bosons ($n_b$) with open boundary conditions (top half), no additional non-3 coloring ground 
states were found. The kagome clusters had completed triangles, resembling their periodic counterparts. 
For kagome clusters on tori (bottom half), additional ground states are found at some low fillings.}
\label{tab:clusters}
\end{center}
\end{table}

\section{36d cluster}
\begin{figure}
\centering
\includegraphics[width=0.90\linewidth]{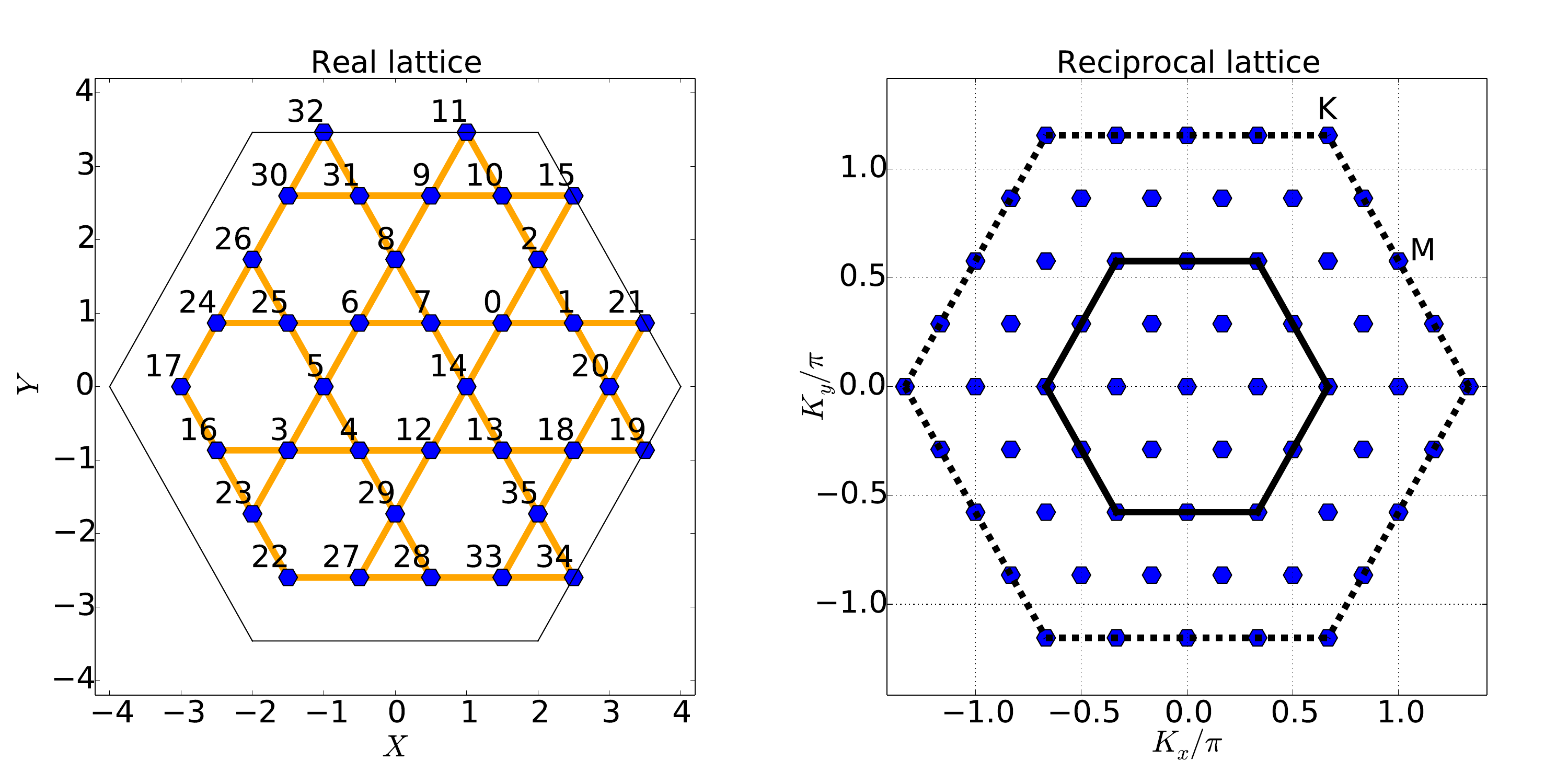}
\caption{(Color online): 36d simulation cluster~(left) and its reciprocal lattice~(right) used in the 
computation of the extended phase diagram. The solid and the dotted lines in reciprocal space show the first and the extended Brillouin zones 
respectively. The high symmetry points, $K$ and $M$, of the latter have been indicated. 
}
\label{fig:36d} 
\end{figure}

Our results for the kagome phase diagram were based on extensive ED calculations on finite lattices. Since ED is severely limited by size 
restrictions, it is important to base our conclusions on simulations of a finite cluster which best represents 
the thermodynamic limit (TDL). The smallest unit cell that can accommodate 
energetically competitive phases, such as the $q=0$ and $\sqrt{3} \times \sqrt{3}$ phases, 
is known to be the 36d cluster, which has been studied by several authors~\cite{Elser93,SudanED} 
focused on exploring the Heisenberg point of the $XXZ$ model i.e. $J_z = 1$.  
This cluster has D6 as its point group symmetry, which includes reflections and 60 degree rotations.
For completeness, in Fig.~\ref{fig:36d}, we show the real space picture of the 36d cluster, along 
with its reciprocal space. 

We work in a fully symmetrized basis which reduces the dimensionality of the Hilbert space for 
a fully symmetric sector to 63044766 basis elements, which is approximately a factor of 144 smaller 
than the original $S_{z}=0$ sector. Although the ground state can belong to any irreducible representation and any momentum sector, by analyzing all of the sectors 
at points $(J_{z},J_{2})$ = $\{(-0.3, \pm 0.05), (0,\pm0.05), (0.1,0), (0.5,\pm 0.02)\}$ 
we conclude that it resides in the symmetric sector of $K=(0,0)$ in the range of interest and focus on investigation of this sector. 

\begin{figure}
\centering
\includegraphics[width=0.90\linewidth]{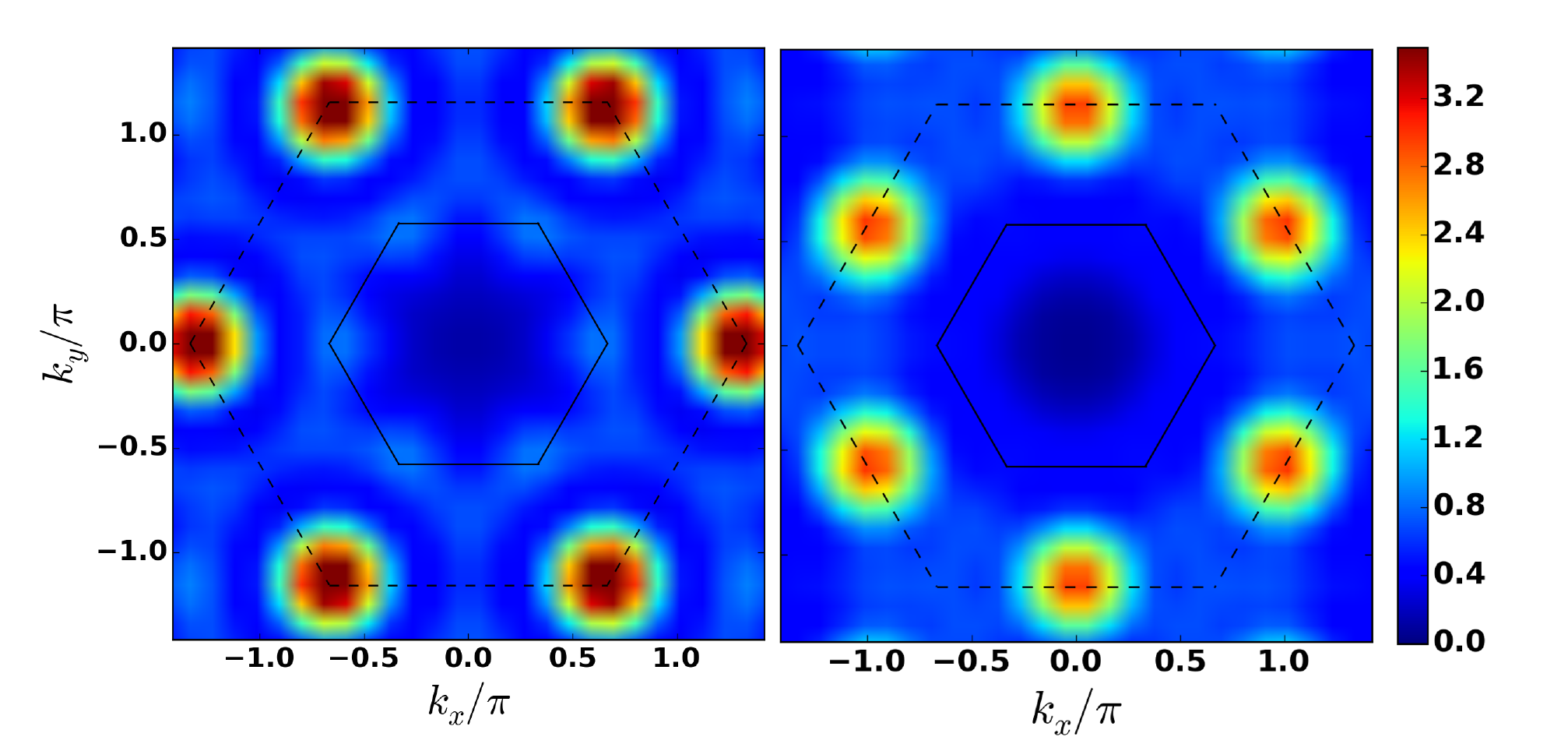}
\caption{(Color online):  Structure factors for magnetically ordered phases
$\sqrt{3} \times \sqrt{3}$ (left) and $q=0$ (right) computed from Exact Diagonalization at $J_{z}=0.1$ $J_{2}=-0.1$ and $J_{z}=0.1$ $J_{2}=0.1$ respectively. For $J_2<0$, the intensity is maximum at the 
$K$ points of the extended Brillouin zone and for $J_2>0$ it is maximum at the $M$ points. }
\label{fig:strucutreFactors} 
\end{figure}

We extract several physical quantities from the ground state vectors, 
such as spin-spin correlation, spin structure factors and ground state fidelity.  
As an example, the structure factors of the magnetically ordered 
$q=0$ and $\sqrt{3} \times \sqrt{3}$ are presented in Fig.~\ref{fig:strucutreFactors}.

\section{Fidelity profiles for $J_{2}$ scans}
In Fig. 3 of the main text, we showed the phase diagram for the kagome antiferromagnet 
in the parameter space of $J_z$ and $J_2$, for the model Hamiltonian,
\begin{eqnarray}
	H[J_z,J_2] &=& \Big( \sum_{\langle i,j \rangle} S^{x}_i S^{x}_j +S^{y}_i S^{y}_j + J_z S^{z}_i S^{z}_j \Big) + J_2 \Big(\sum_{\langle \langle i,j \rangle \rangle} S^{x}_i S^{x}_j +S^{y}_i S^{y}_j + J_z S^{z}_i S^{z}_j \Big) \nonumber \\
		   &=& H^\textrm{nn}_{XXZ}[J_z] + J_2 H^\textrm{nnn}_{XXZ}[J_z]
\end{eqnarray}
where $\langle i,j \rangle$ and $\langle \langle i,j \rangle\rangle$ denote the nearest neighbor (nn) and next-nearest-neighbor (nnn) sites respectively. 
 
Our estimates of the phase boundaries were based on the measuring fidelity of the ground state wavefunction 
$\psi (J_z, J_2)$, by scanning in the $J_2$ direction (keeping $J_z$ fixed),
\begin{equation}
f(J_z, J_2) \equiv \Big| \langle \psi (J_z, J_2 - \Delta J_2/2 )|\psi (J_z,J_2 + \Delta J_2/2) \rangle \Big|
\end{equation}
where $\Delta J_2$ is the step size. Dips in the fidelity profile indicate the existence of phase transitions. 

\begin{figure}[htpb]
\centering
\includegraphics[width=0.77\linewidth]
{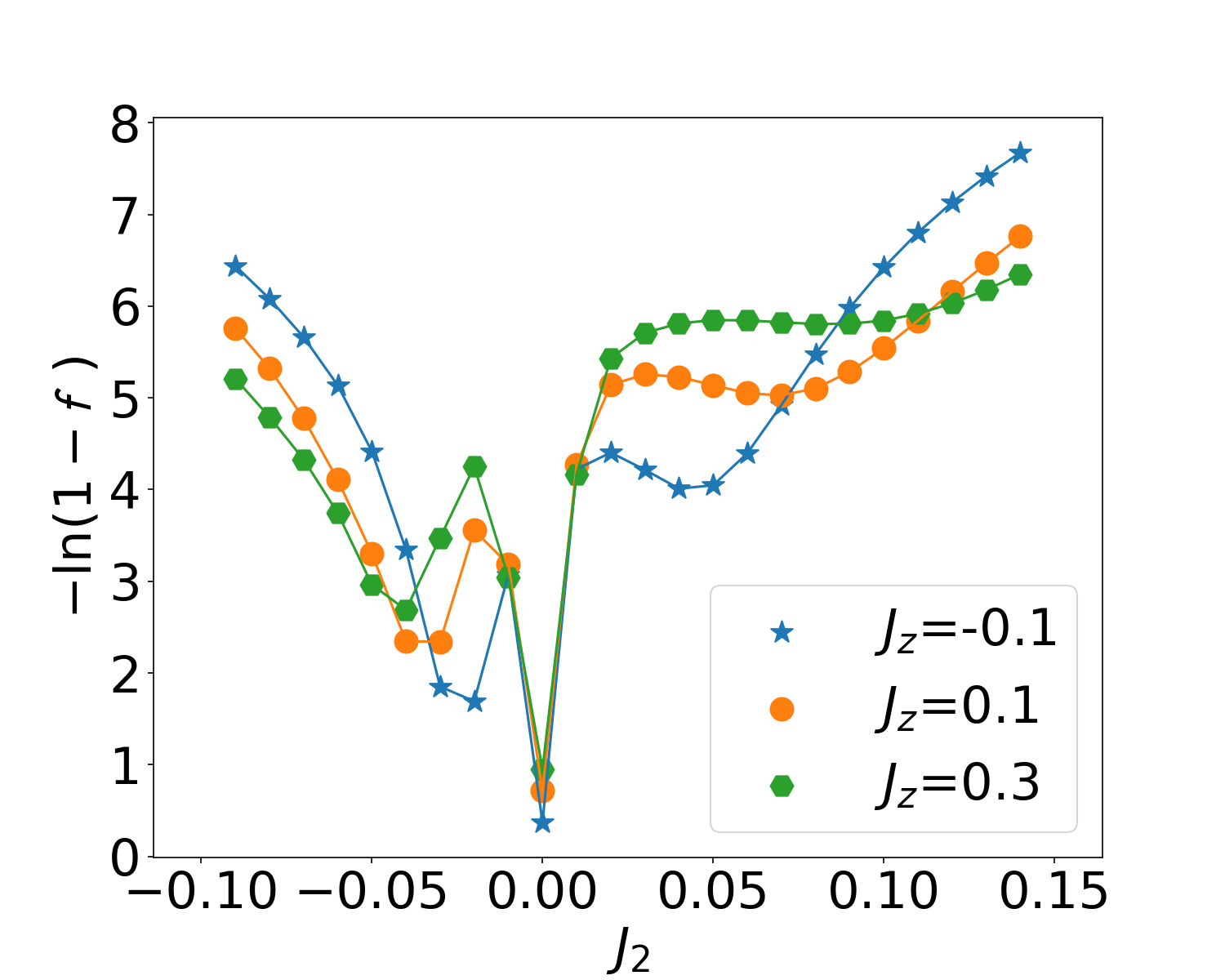}
\caption{ (Color online) Fidelity scans along the $J_{2}$ axis for different values of $J_{z}$. 
The fidelity is evaluated for ground state wavefunctions at parameter values 
which differ by $\Delta J_2=0.01$ (keeping $J_z$ fixed). 
The spreading location of the leftmost and rightmost dips (increase in $|J_2|$), 
as $J_z$ is increased, corresponds to the wedge feature in the kagome phase diagram. } 
\label{fig:fidelity}
\end{figure}

Our results for representative $J_z$, with $\Delta J_2 = 0.01$ are shown in Fig.~\ref{fig:fidelity}. We observe that 
there are prominent dips for $J_2<0$ and $J_2 \approx 0$ and only a marginal one for $J_2 > 0$. 
The location of of both the leftmost and rightmost dips increases in $|J_2|$ 
on increasing $J_z$, this corresponds to the appearance of the wedge in the 
kagome phase diagram in Fig. 3. Prominently, the dip at $J_2 \approx 0$ is present for all 
$J_z$ shown.
 
\section{Ferromagnet and boundaries shared with adjoining phases}
In this section of the supplement, we discuss certain aspects of the ferromagnetic (FM) region 
reported in Fig.~3 of the phase diagram and the phase boundaries it shares with its adjoining phases. 

Moving along the direction of $J_{z} < -0.5$ lifts the exponential 
degeneracy to favor the fully polarized sector. Therefore, in the $S_z=0$ sector, the energy density (energy per site) 
is minimized by the phase separated FM state which has half the system ($N$ sites) maximally polarized up ($S_z=N/2$) 
and the other half maximally polarized down ($S_z=-N/2$).  This can be proven analytically, 
since the fully polarized state simultaneously generates the minimal possible energy for all 
four terms of the Hamiltonian 
($XXZ0$ of the triangles made of nearest neighbor bonds, $J_2 XXZ0$ of the triangles 
made of next nearest neighbor bonds, $(J_z+1/2) \sum_{\langle i,j \rangle} S_i^{z} S_j^{z}$ 
and $J_2 (J_z + 1/2) \sum_{\langle\langle i,j \rangle\rangle} S_i^{z}S_j^{z}$) for $J_2 \ge 0$. 
Phase separation results in a domain wall which costs absolute energy but, in the TDL, 
costs zero energy per site for short-ranged Hamiltonians such as ours.  
While it is possible there are other states with the same energy density but lower absolute energy for a finite system, 
we see no evidence of this. 
\begin{figure}
\centering
\includegraphics[width=0.85\linewidth]{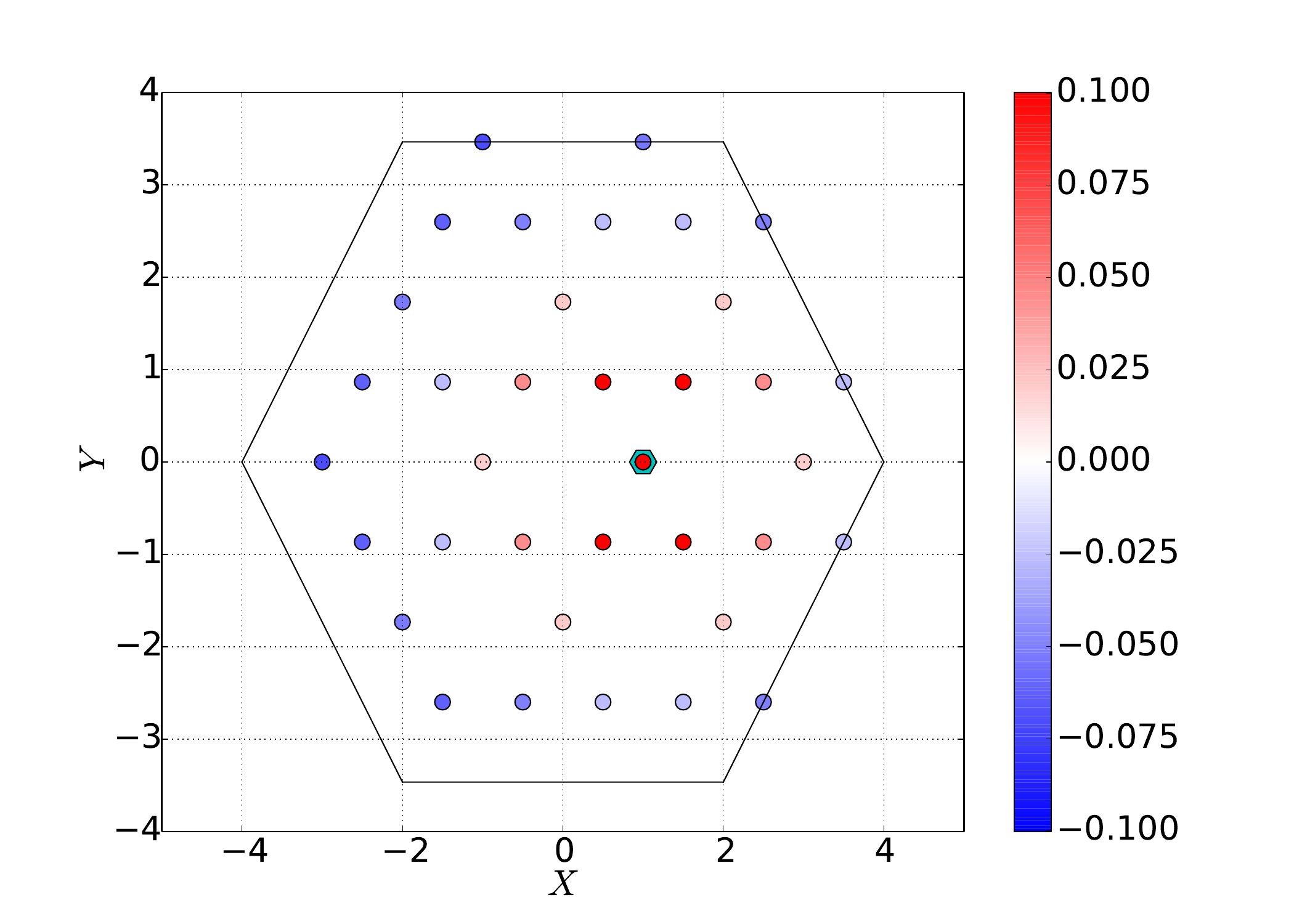}
\caption{(Color online): ZZ component of the real space spin-spin correlation function (with respect to a reference site) 
in the $S_{z}=0$ ground state at $J_{z}=-0.7$, $J_{2}=0$. The cyan hexagon marks the lattice site with respect to 
which the correlation function is computed. The color represents the correlation strength 
(red - ferromagnetic, blue antiferromagnetic correlations).}
\label{fig:phaseSeparation} 
\end{figure}
While a large enough simulation will exhibit emergent phase separation, 
finite size effects dominate in a small ED calculation. Nonetheless, in most (but not all) 
of the region $J_z<-0.5, J_2>0$  we see clear phase separation in the spin-spin correlation function such as at $J_{z}=-0.7$, $J_{2}=0.0$, 
see Fig. \ref{fig:phaseSeparation}.  

Let us now consider the lines separating the $q=0$ and FM (the vertical line $J_z=-1/2$ for $J_2 \geq 0$) 
and the $\sqrt{3}\times\sqrt{3}$ and FM regions, the latter calculated to be,
\begin{equation}
|J_z| = \frac{(\frac{1}{2} + |J_2|)}{(1 - |J_2|)}
\label{eq:JzJ2}
\end{equation}
Both boundaries can be understood by comparing the semiclassical energy of the unprojected magnetically 
ordered states with that of the FM. 
For example, the energy associated with four nearest neighbor and four next nearest neighbor bonds emanating 
from a single site in the FM state is $- 4 |J_z| + 4|J_2||J_z|$ in comparison to $4 (-1/2)-4 |J_2|$ 
for the unprojected coplanar $\sqrt{3} \times \sqrt{3}$ state. The phase boundary of these two phases 
is shown by the solid line in Fig.~3 of the main text 
and corresponds to Eq.~\eqref{eq:JzJ2}. Similarly, the $q=0$ energy is higher than the FM for $J_z<-1/2$ for any $J_2 > 0 $. 
We note that despite involving only semiclassical arguments, 
the agreement of these phase boundary estimates with those obtained from 
energy densities calculated from ED, is excellent. 

\section{Hamiltonian with four coloring exact ground states}
We noted that the idea of coloring wavefunctions generally applies to beyond triangular motifs. 
Here we explicitly write down the Hamiltonian for which the four coloring 
wavefunction is an exact ground state on lattices with motifs involving 
four sites (such as the square, checkerboard and pyrochlore lattices).  
We derive this Hamiltonian for the case of four sites; the extension 
to the case of lattices with \emph{shared} four colorable motifs is trivial. 

First, define the four colors as,
\begin{subequations}
\begin{eqnarray}
	| a \rangle &\equiv& | \uparrow \rangle +  | \downarrow \rangle \\
	| b \rangle &\equiv& | \uparrow \rangle + i| \downarrow \rangle \\
	| c \rangle &\equiv& | \uparrow \rangle -  | \downarrow \rangle \\
	| d \rangle &\equiv& | \uparrow \rangle - i| \downarrow \rangle
\end{eqnarray} 
\end{subequations}

Then define the states, 
\begin{subequations}
\begin{eqnarray}
	| 1 \rangle &\equiv& | \downarrow \uparrow \uparrow \uparrow \rangle + | \uparrow \downarrow \uparrow \uparrow \rangle + | \uparrow \uparrow \downarrow \uparrow \rangle + | \uparrow \uparrow \uparrow \downarrow \rangle \\
	| 2 \rangle &\equiv& | \uparrow \uparrow \downarrow \downarrow \rangle + | \uparrow \downarrow \uparrow \downarrow \rangle + | \uparrow \downarrow \downarrow \uparrow \rangle + | \downarrow \uparrow \uparrow \downarrow \rangle + | \downarrow \uparrow \downarrow \uparrow \rangle + | \downarrow \downarrow \uparrow \uparrow \rangle \\
	| 3 \rangle &\equiv& | \uparrow \downarrow \downarrow \downarrow \rangle + | \downarrow \uparrow \downarrow \downarrow \rangle + | \downarrow \downarrow \uparrow \downarrow \rangle + | \downarrow \downarrow \downarrow \uparrow \rangle
\end{eqnarray}
\end{subequations}

Then, any Hamiltonian of the form
\begin{equation}
	H = \lambda_1 |1 \rangle \langle 1 | + \lambda_2 |2 \rangle \langle 2 | + \lambda_3 |3 \rangle \langle 3 |
\end{equation}
with $\lambda_1, \lambda_2, \lambda_3 \ge 0$ will have the coloring wavefunction $|C \rangle = | a \rangle \otimes | b \rangle .... $
as an exact ground state with zero energy as long as one satisfies the constraint of one $a$, $b$, $c$, $d$ each per four-site 
motif. Here we present the result for $\lambda_1=\lambda_2=\lambda_3=1$, where $H$ is also time reversal invariant.

We used the DiracQ package~\cite{DiracQ} to simplify the spin algebra and up to an overall scale factor 
found the Hamiltonian to be, 
\begin{eqnarray}
H & = & \frac{7}{8} + \Big( \sum_{i<j} S_i^{+}S^{-}_j + S^{-}_i S^{+}_j - \frac{1}{2} S^{z}_{i} S^{z}_{j} \Big) + \sum_{i<j, k<l, \text{diff}}  S_i^{+}S^{+}_j  S_k^{-}S^{-}_l  - 2\; S^{z}_{1} S^{z}_{2} S^{z}_{3} S^{z}_{4} 
\end{eqnarray}
where the notation "diff" is used to indicate that all the indices $i,j,k,l$ are distinct. 

\end{document}